\documentclass[aps,twocolumn,prb,showpacs,superscriptaddress]{revtex4-2}
\usepackage{makeidx}
\usepackage{amsfonts}
\usepackage{graphics}
\usepackage{graphicx}
\usepackage{amssymb}
\usepackage{amsthm}
\usepackage{amsmath}
\usepackage{hyperref}

\begin{document}

\title{A note on an inversion algorithm for vertical ionograms for the prediction of plasma frequency profiles}

\author{R. Kenyi Takagui-Perez}
\affiliation{Pontifical Catholic University of Peru}


\begin{abstract}
    Building upon the concept of utilizing quasi-parabolic approximations to determine plasma frequency profiles from ionograms, we present a refined multi-quasi-parabolic method for modeling the E and F layers. While a recent study [AIP Advances 14, 065034 (2024)] \cite{niu2024} introduced an approach in this direction, we identified several inaccuracies in its mathematical treatment and numerical results. By addressing these issues, we offer a clearer exposition and a more robust algorithm. Our method assumes a parabolic profile for the E layer and approximates the F layer with a series of concatenated quasi-parabolic segments, ensuring continuity and smoothness by matching derivatives at the junctions. Applied to daylight ionograms from the Jicamarca Observatory in Lima, our inversion algorithm demonstrates excellent agreement between the synthetic ionograms generated from our predicted plasma frequency profiles and the original measured data.
\end{abstract}

\maketitle


\section{Introduction}

\label{intro}

Most of our knowledge about the ionosphere comes from ionogram records. These $h'(f)$ records give the apparent or virtual heights of reflection $h'(f)$ of a vertically transmitted radio wave, as a function of the wave frequency $f$. This paper aims to retrieve the electronic density profile from measured ionograms.

The analysis of ionograms consists basically of converting an observed $h'(f)$ curve, which gives the virtual height of reflection $h'$ as a function of the wave frequency $f$, into an $N(h)$ curve giving the variation of the electron density $N$ with height $h$. These two curves are related by
\begin{equation}
h'(f) = \int_0^{h_r} \mu' \, dh,
\label{eq:first}
\end{equation}
where the group refractive index $\mu'$ is a complicated function of $f$, $N$, and the strength and direction of the magnetic field. The height of reflection, $h_r$ for the frequency $f$ depends on $f$, $N$, and (for the extraordinary rays only) the strength of the magnetic field.

Previous efforts to solve this ill-posed problem have included lamination techniques, least-squares polynomial approximations, and ray tracing methods. Since there is no analytic solution to Eq.~(\ref{eq:first})---that is, no direct expression for \( N(h) \) in terms of \( h'(f) \)---researchers have explored various numerical and approximation strategies. The lamination technique proposed by Reilly~\cite{reilly1989} involves assuming various \( N(h) \) model curves, passing them through a forward model to generate corresponding ionograms, and then comparing the resulting \( h'(f) \) curves with those observed experimentally. In contrast, Reinisch et al.~\cite{reinisch1983,titheridge1967} utilized Chebyshev polynomial methods to approximate the F layer, aiming for a more efficient process in ionogram analysis. However, their software is proprietary, making it difficult to replicate and improve upon their work. More recently, Ankita et al.~\cite{ankita} introduced a different approach by using electromagnetic wave propagation simulations based on Hamiltonian formulations for ray tracing.

In this study, we build upon the concepts introduced by Niu et al. \cite{niu2024}, who proposed using multivariate quasi-parabolic layers to develop an inversion algorithm for approximating the plasma frequency profiles of the E and F ionospheric layers. While their underlying idea holds potential, we have identified several mathematical errors, mislabeled equations, and inconsistencies in their manuscript that make the methodology challenging to follow and replicate. To address these issues, we offer a clearer and more accurate exposition of the intended approach, providing detailed computations for transparency and understanding. This work is intended to be pedagogical, aiming to enhance comprehension of the inversion process. Additionally, recognizing the scarcity of open-source software for ionogram inversion, we are releasing our code to the community to facilitate further research and application.

An introduction for the $QP$ layers and a treatment of how to use them to reconstruct the electron density profile is presented in Sec. \ref{sec:intro_QP}. In Sec. \ref{sec:algorithm} we present a detailed inversion algorithm using ideas from the previous section. In Sec. \ref{sec:forward_model} we describe the forward model, a series of detailed calculations to obtain the virtual heights given a plasma frequency profile. Finally, in Sec. \ref{sec:results} we show some results of the predicted profiles based on a given ionograms. 

\section{Model: QP layers}
\label{sec:intro_QP}
The quasi-parabolic model of the ionosphere is useful for representing electron density variations in layers such as the E and F regions; it was introduced by Forsterling et al. \cite{forsterling1931}. The electron density $N_e$ as a function of the radial distance $r$ from the Earth's center is described by:

\[
N_e =
\begin{cases}
N_m \left[ 1 - \left( \frac{r - r_m}{y_m} \right)^2 \right], & \text{for } r_b < r < r_m + y_m \\
0, & \text{otherwise}
\end{cases}
\]

where:

\begin{itemize}
    \item $N_e$: Electron density, which attains a maximum value $N_m$ at the center of the ionospheric layer.
    \item $r$: Radial distance from the Earth's center.
    \item $r_m$: Radial distance where the electron density reaches its peak value $N_m$.
    \item $r_b$: Radial distance at the base of the ionospheric layer, defined as $r_m - y_m$.
    \item $y_m$: Half-thickness of the ionospheric layer, controlling the layer's vertical extent.
    \item $N_m$: Maximum electron density at $r = r_m$.
\end{itemize}

This model confines the electron density within a parabolic profile, smoothly decaying to zero outside the bounds $r_b < r < r_m + y_m$. A very slight modification to the parabolic model permits the derivation of exact equations for ray-path parameters. This modified parabolic ionosphere will be termed the "quasi-parabolic" or, more simply, the "QP" model.

\begin{figure}[hb]
    \begin{center}
    \includegraphics*[width=0.90\columnwidth]{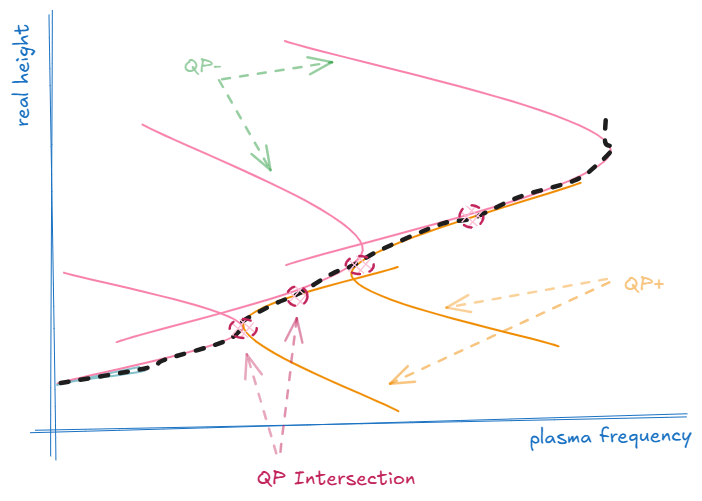}
    \end{center}
    \caption{(Color online) Description of what the concatenation of QP layers aims to build in a plasma frequency vs real height plane. The dotted black line would be the plasma frequency profile.}
    \label{fig:cute}
\end{figure}

We make a $QP$ layer the basic unit that will be used to model split sections of the F layer and the entirety of the E layer. We define a $QP$ layer as

\begin{equation}
f_{ni}^2 = f_{ci}^2 \left[ 1 \pm \left( \frac{r - r_{mi}}{y_{mi}} \right)^2 \left( \frac{b_i}{r} \right)^2 \right]
\label{eq:qp_eq}
\end{equation}
where $f_{ni}$ is the plasma frequency dependent on the real height $r$ and $f_{ci}$ is the critical frequency. However these physical meanings are lost once we consider the $QP$ layer as a unit, and instead it is more a geometric construction.
It can also be written as

\begin{equation}
y_i = f_{ni}^2 = a_i \pm b_i \left( 1 - \frac{r_{mi}}{r} \right)^2
\label{eq:QP_other_form}
\end{equation}

where $y_i = f_{ni}^2$, $a_i = f_{ci}^2$, and $b_i = a_i \left( \frac{b_i}{y_{mi}} \right)^2$. There will be two kinds of QP layers: $QP_i^+$ and $QP_i^-$. Therefore they will be represented by equations $y_i^+$ and $y_i^-$, correspondingly. Each QP layer is parameterized by three numbers: $a_i$, $b_i$, and $r_{mi}$. See Fig \ref{fig:cute}.

The way we use Eq. (\ref{eq:qp_eq}) to find the real heights, is solving the equation for a given frequency. That is, per QP layer, we select a bunch of plasma frequency points and solve the equation getting their corresponding real heights.  

\subsection{Data}

We assume we have the autoscaled ionogram virtual heights for frequencies $f_i \in [0, f_F ]$ with arbitrary strides between frequency points, and also the frequency position of the E layer
critical frequency $f_E$. Big gaps between frequency data points may cause undesired behaviour in the algorithm.

\subsection{E layer}

It is well known that the E-layer virtual heights per frequency $f$ can be modeled as

\begin{equation}
h'(f) = r_b + \frac{1}{2} y_m \frac{f}{f_E} \ln \left( \frac{f_E + f}{f_E - f} \right)
\label{eq:e_layer}
\end{equation}

Where $r_b$ is the height where the ionogram starts. In the case of E layers, we usually don't have the complete trace so it is necessary to search for $r_b$ and $y_m$ values. If we were to use a $QP$ model for the E layer instead of Eq. (\ref{eq:e_layer}), we would still need to compute the same values since we take only $f_E$, the E layer critical frequency, for granted. We use a brute-force approach to find the best parameters for the E-layer using a simple two for-loop. For each pair $(r_{bE}, y_{mE})$ we compare the produced virtual heights $h'(f)$ using the Eq. (\ref{eq:e_layer}), or with the forward model in case we use a $QP$ model, versus the original ionogram. Finally, the probe pair with the least error difference is the one we take. The metric used for measuring the error we use is the root-mean-squared error.

Because the E layer presents a pronounced steep, we consider a probe frequency points that are not evenly distributed in, say, a range $[A,B]$, but more densily distributed near the upper bound $B$. We use the following equation to obtain a suitable distribution of values
\begin{equation}
    x_i = A + (B - A) \frac{1 - \exp\left(-k \frac{i}{N}\right)}{1 - \exp(-k)}.
\end{equation}  

\subsection{QP layers concatenation for F layer reconstruction}
Next, we approximate the F layer by concatenating several $QP$ layers by alternating between $QP^-$ to $QP^+$ and $QP^+$ to $QP^-$ layers. We start the F layer with a $QP^+$ layer. We ensure the continuity of the plasma frequency profile between $QP_i$ and $QP_{i-1}$ by making its derivatives equal at $r_i$, the real height corresponding to last frequency point we considered in the $QP_{i-1}$ layer. See Fig. \ref{fig:cute}.

\begin{equation}
\left. \frac{dy_i^\pm}{dr} \right|_{r=r_i} = \left. \frac{dy_{i-1}^\mp}{dr} \right|_{r=r_i} \quad \text{or} \quad \left. \frac{dy_i^\mp}{dr} \right|_{r=r_i} = \left. \frac{dy_{i-1}^\pm}{dr} \right|_{r=r_i},
\end{equation}

From these relations we can get expressions for $b_i$ and $r_{mi}$, dependent on $a_i$ and the parameters of the $QP_{i-1}$ layer. See Eq. (\ref{eq:QP_other_form})

Let us compute the derivatives for the different types of $QP$ layers. For $y_i^-$

\begin{equation}
y_i^-(r) = a_i - b_i \left( 1 - \frac{r_{mi}}{r} \right)^2
\end{equation}

its corresponding derivative is

\begin{equation}
y_i^{-}(r)' = -\frac{2b_i r_{mi}}{r^2} \left( 1 - \frac{r_{mi}}{r} \right)
\end{equation}

On the other hand, for $y^+$

\begin{equation}
y_i^+(r) = a_i + b_i \left( 1 - \frac{r_{mi}}{r} \right)^2
\end{equation}

its derivative is

\begin{equation}
y_i^{+}(r)' = \frac{2b_i r_{mi}}{r^2} \left( 1 - \frac{r_{mi}}{r} \right)
\end{equation}

Then, as we mentioned, there will be two cases: $QP^- \text{ to } QP^+$ or $QP^+ \text{ to } QP^-$. First, for the case of $QP_{i-1}^-$ to $QP_i^+$, to find the dependence of parameters $b_i$ and $r_{mi}$ with the ones of $QP_{i-1}$ and ensure the continuity of the curve, we equate $y_{i-1}^{-}(r_i)' = y_i^{+}(r_i)'$. $r_i$ is the point where both curves meet, computationally speaking $r_i$ is the last point we took from the curve $QP_{i-1}^-$. We find

\begin{equation}
r_{mi} = \frac{r_i^2 y_{i-1}^{-}(r_i)'}{2(y_{i-1}(r_i) - a_i) + r_i y_{i-1}^{-}(r_i)'}
\end{equation}

\begin{equation}
b_i = \frac{\left[ 2(y_{i-1}(r_i) - a_i) + r_i y_{i-1}^{-}(r_i)' \right]^2}{4(y_{i-1}(r_i) - a_i)}
\end{equation}

Finally, for the case of $QP_{i-1}^+$ to $QP_i^-$, we equate

\begin{equation}
y_{i-1}^{+}(r_i)' = y_i^{-}(r_i)'.
\end{equation}

\begin{equation}
r_{mi} = \frac{r_i^2 y_{i-1}^{+}(r_i)'}{2(y_{i-1}^+(r_i) - a_i) + r_i y_{i-1}^{+}(r_i)'}
\end{equation}

\begin{equation}
b_i = -\frac{\left[ 2(y_{i-1}^+(r_i) - a_i) + r_i y_{i-1}^{+}(r_i)' \right]^2}{4(y_{i-1}^+(r_i) - a_i)}
\end{equation}

\section{Algorithm}
\label{sec:algorithm}

The general idea is that for each $QP$ layer we will try different values of $f_{ci}$, each giving a its own set of parameters, which at the same time depend on values computed in the previous $QP_{i-1}$. Next, for each $f_{ci}$, we use a batch of $N$ plasma frequency points for which we calculate their real heights and subsequently their virtual heights using the forward model. We compare them against the measured virtual heights and keep the batch with the least error. Finally, we append the best one and move on onto the next layer. 

A more detailed explanation is as follows: the algorithm starts with the calculation of the E-layer, also known as $QP^-_0$ layer. Then, we can iteratively compute the next $QP_i$ parameters since we already know the ones from the $QP_{i-1}$ layer meaning $r_{mi}$ and $b_i$ will be left as a function of $a_i = f_{ci}^2$. To find $f_{ci}$, we use an exhaustive search algorithm, which means trying several values of $f_{ci}$ by brute force. If the $QP_i$ layer is a $QP^+$ layer, we will search $f_{ci}$ in the range $[f_L + \epsilon, f_L + 2.0]$, and, if, on the other hand, we are computing a $QP^-$ layer, the range will be $[f_L - 2.0, f_L - \epsilon]$. For each of these $f_{ci}$ values, we take a batch of $N$ plasma frequency points including $f_{ci}$ and compute their real and virtual heights with the help of the forward mode and compare them against the real measurements from the ionogram. The batch producing the least error is attached to our \texttt{fp profile} array. Once we are done with the $QP_i$ layer, we consider the last element of the \texttt{fp profile} to be the intersection point between the $QP_i$ and $QP_{i+1}$ layers. Finally, we repeat the process. The number of $QP$ layers should be defined beforehand.

\section{Forward Model}
\label{sec:forward_model}

Each time we add new plasma frequency and real height data points to our plasma frequency profile curve, we need to test it against the original ionogram and calculate the error. To achieve this we need a forward model that can calculate an ionogram given a plasma frequency profile.

\begin{figure}[t]
    \begin{center}
    \includegraphics*[width=0.70\columnwidth]{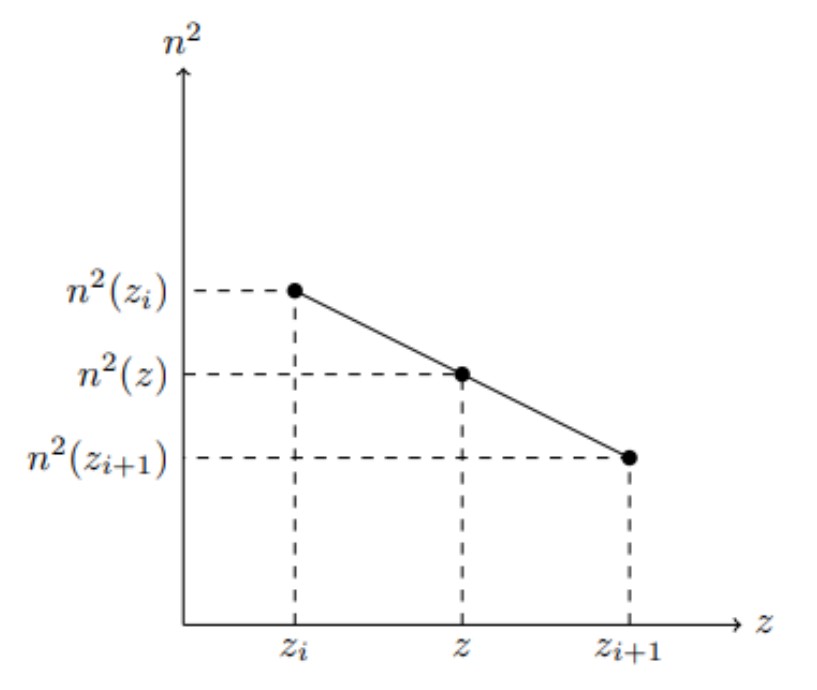}
    \end{center}
    \caption{Close up to the $n(z)^2$ function.}
    \label{fig:grafiquito}
\end{figure}

Another way to express the ionospheric virtual height reflection is
\begin{equation}
h(f) = \int_0^{z_r} \frac{dz}{n(z)}
\label{eq:virtual_height}
\end{equation}
and by looking at the expression of \(n(z) = \sqrt{1 - \frac{f_p(z)^2}{f^2}}\), plus the condition that \(f_p(z_r) = f\), we can foresee that the curve of \(n(z)^2\) will look like an inverse function beginning in the coordinate \((0, 1)\) and then converging towards \(n(z)^2 = 0\) at \(z = z_r\). If we zoom in on a curve segment, the ending points would be like the ones in Fig. \ref{fig:grafiquito}. Using triangle relations we get
\begin{equation}
\frac{z_{i+1} - z_i}{n^2(z_i) - n^2(z_{i+1})} = \frac{z - z_i}{n^2(z_i) - n^2(z)}
\end{equation}

where \(n(z)\) can be extracted from this equation
\begin{equation}
n^2(z) = n^2(z_i) - (z - z_i) \left[\frac{n^2(z_i) - n^2(z_{i+1})}{z_{i+1} - z_i}\right]
\end{equation}

and replace it in Eq. \ref{eq:virtual_height}. Because the integrand is an inverse function the biggest contributions will come from the segment with \(z_i\) value closer to \(z_r\). If we treat the integral from Eq. \ref{eq:virtual_height} by parts (imagine our line of integration is formed by \(N\) points), then there will be two different contributions, one coming from the last segment \((z_1, z_r)\) and another from \((z_0, z_1)\).
\begin{equation}
h(f) = \int_{z_1}^{z_r} \frac{dz}{n(z)} + \int_0^{z_1} \frac{dz}{n(z)}
\label{eq:two_integrals}
\end{equation}

First, we deal with the latter.
\begin{equation}
    \begin{split}
        &= \sum_{i=0}^{N-2} \int_{z_i}^{z_{i+1}} \frac{dz}{n(z)}\\
        &= \sum_{i=0}^{N-2} \int_{z_i}^{z_{i+1}} \frac{1}{\sqrt{n^2(z_i) - (z - z_i) \left[\frac{n^2(z_i) - n^2(z_{i+1})}{z_{i+1} - z_i}\right]}} \, dz
    \end{split}
\end{equation}

Doing a change of variables \(z' = z - z_i\) we are left with
\begin{equation}
    \begin{split}
        &= \int_0^{z_1} \frac{dz}{n(z)}\\
        &= \sum_{i=0}^{N-2} \int_0^{z_{i+1} - z_i} \frac{1}{\sqrt{n^2(z_i) - z' \left[\frac{n^2(z_i) - n^2(z_{i+1})}{z_{i+1} - z_i}\right]}} \, dz',
    \end{split}
\end{equation}

Then using the result \(\int \frac{1}{\sqrt{a - b x}} dx = -\frac{2\sqrt{a - b x}}{b} + C\) to calculate the integral analytically
\begin{equation}
\begin{split}
    \int_0^{z_1} \frac{dz}{n(z)} &= \sum_{i=0}^{N-2} \frac{2 (z_{i+1} - z_i)}{\sqrt{n^2(z_i)} + \sqrt{n^2(z_{i+1})}}\\
    &= \sum_{i=0}^{N-2} \frac{2 \Delta z}{n(z_i) + n(z_{i+1})}
\end{split}
\end{equation}

Secondly, we deal with the first integral in Eq. (\ref{eq:two_integrals}). As the upper limit for the integral is $z_r$, we need to remember that $n(z = z_r) = 0$, then because of 
\(
z_r = \frac{n^2(z_i)}{n^2(z_i) - n^2(z_{i+1})} (z_{i+1} - z_i) + z_i
\)
the integral will be
\begin{equation}
n^2(z_i) \int_0^{\frac{z_{i+1} - z_i}{n^2(z_i) - n^2(z_{i+1})}} \frac{dz}{n(z')}
\end{equation}
already applied the change of variables. The result is
\begin{equation}
\int_{z_i}^{z_r} \frac{dz}{n(z)} = 2n(z_i) \frac{(z_{i+1} - z_i)}{n^2(z_i) - n^2(z_{i+1})}
\end{equation}
with $z_i = z_{N-1}$ and $z_{i+1} = z_N$. Then, by calculating both analytically both integrals we can get an accurate value for the virtual height given a plasma frequency profile.

\section{Results}
\label{sec:results}

To test our inverse algorithm and forward model, we used data measured in the Jicamarca Observatory in Lima, Peru. We concentrate in daylight ionograms as they present some data points for the E layer. We do not support ionograms with only the F layer present.

The original measured ionograms will be represented by magenta dots, the reconstructed plasma frequency profile in a yellow solid line, and the synthetic ionogram calculated from the reconstructed plasma frequency profile will be represented by empty black squares. For the sake of brevity, we present results with an hour of difference during daylight.

\begin{figure*}[htbp]
    \begin{center}
    \includegraphics*[width=1\columnwidth]{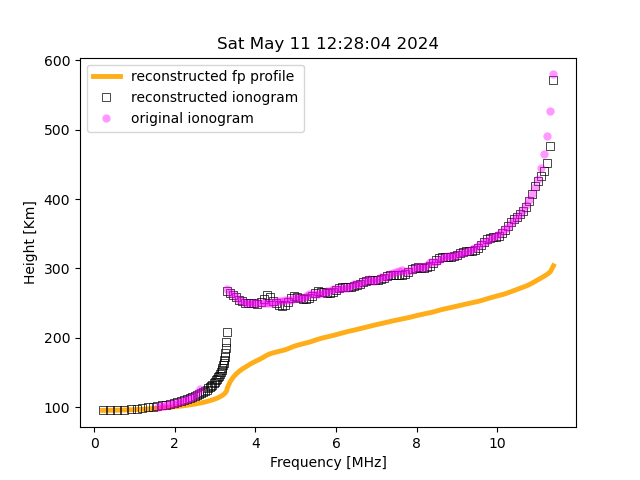}
    \end{center}
    \caption{}
    \label{fig:04}
\end{figure*}

\begin{figure*}[htbp]
    \begin{center}
    \includegraphics*[width=1\columnwidth]{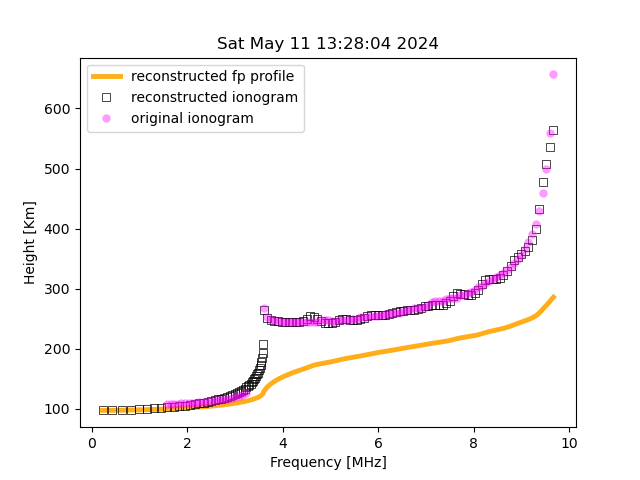}
    \end{center}
    \caption{}
    \label{fig:05}
\end{figure*}

\begin{figure*}[htbp]
    \begin{center}
    \includegraphics*[width=1\columnwidth]{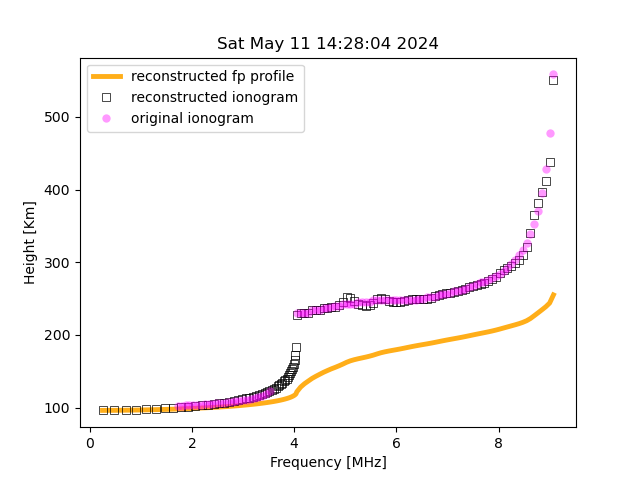}
    \end{center}
    \caption{}
    \label{fig:06}
\end{figure*}

\begin{figure*}[htbp]
    \begin{center}
    \includegraphics*[width=1\columnwidth]{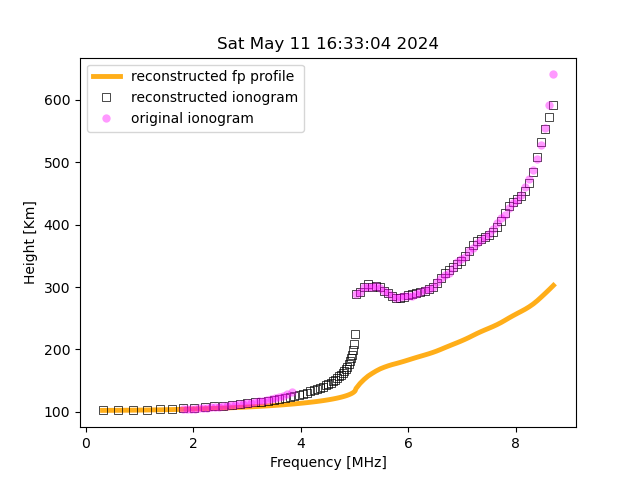}
    \end{center}
    \caption{}
    \label{fig:08}
\end{figure*}

\begin{figure*}[htbp]
    \begin{center}
    \includegraphics*[width=1\columnwidth]{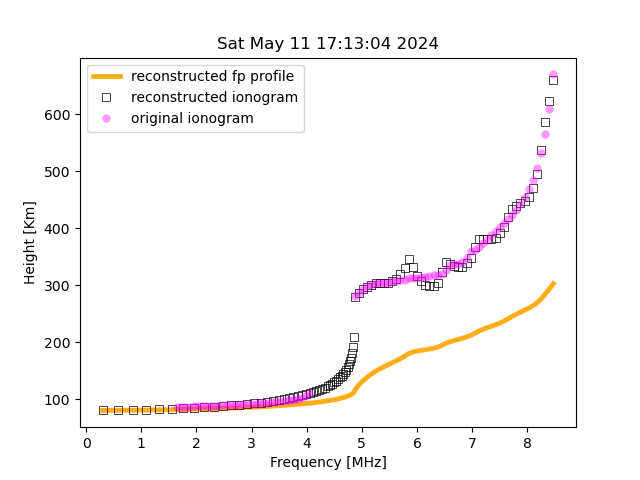}
    \end{center}
    \caption{}
    \label{fig:09}
\end{figure*}

Even though the final reconstructed ionogram does not perfectly fit the original one, see Fig. \ref{fig:09} , in general shows good agreement, see Figs. \ref{fig:04} \ref{fig:05} \ref{fig:06} \ref{fig:08}. But more importantly, the results give us a good idea of what the plasma frequency or electron density profile look like.

To obtain this results, we usually computed between $25$ to $30$ $QP$ layers with $4$ data points per layer. However, to stabilize the virtual heights of the first produced synthetic ionogram data points in the F layer, we used a single data point for the first $10$ $QP$ layers. Additionally, our algorithm is able to complete the missing data points in the E layer, see for example Fig. \ref{fig:09}.

The open-source code and data used for this research can be found in Github repository \href{https://github.com/TAOGenna/inversion-algorithm-plasma-frequency-profile}{\underline{InversionAlgorithm}}.
\begin{figure*}[htbp]
    \begin{center}
    \includegraphics*[width=1\columnwidth]{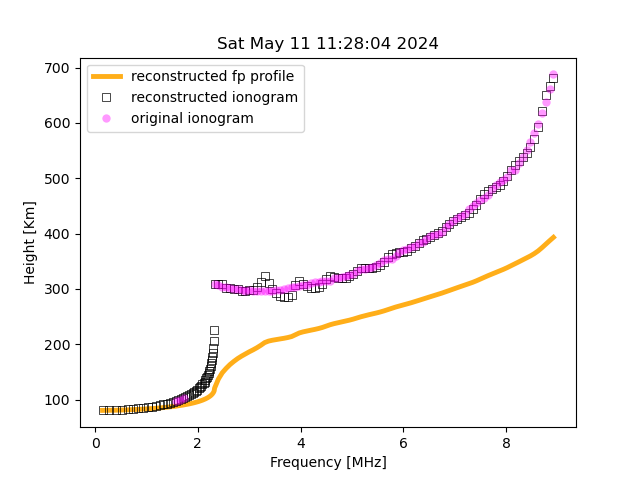}
    \end{center}
    \caption{}
    \label{fig:03}
\end{figure*}

\begin{figure*}[htbp]
    \begin{center}
    \includegraphics*[width=1\columnwidth]{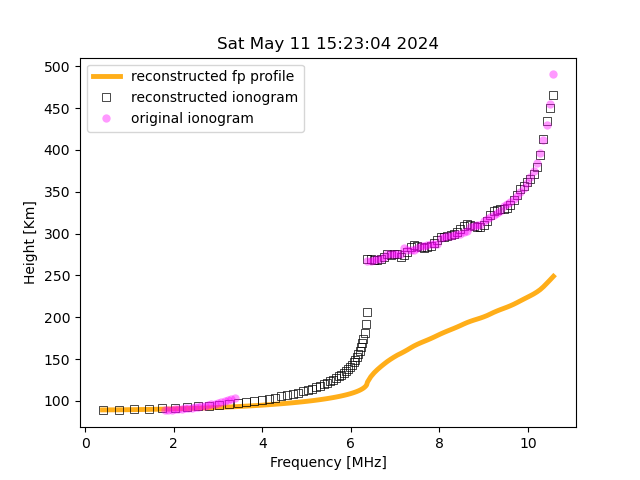}
    \end{center}
    \caption{}
    \label{fig:07}
\end{figure*}

\section*{Acknowledgments}
R. K. T. P. acknowledges the financial support by the Radio Astronomy Institute of the Pontificia Catolica Universidad del Peru and Prof. Marco
Milla for useful comments and discussions.

\end{document}